\documentclass[a4paper,fleqn,usenatbib]{mnras}

\usepackage[T1]{fontenc}
\usepackage{ae,aecompl}

\usepackage{graphicx}	% Including figure files
\usepackage{amsmath}	% Advanced maths commands
\usepackage{amssymb}	% Extra maths symbols

\title[The extra-tidal  features of NGC\,6864]{DECam photometry reveals 
extra-tidal stars around the Milky 
Way globular cluster NGC	\,6864 (M75)}

\author[Andr\'es E. Piatti]{
Andr\'es E. Piatti$^{1,2}$\thanks{E-mail: andres.piatti@unc.edu.ar} \\
$^{1}$Instituto Interdisciplinario de Ciencias B\'asicas (ICB), CONICET-UNCUYO, Padre J. Contreras 1300, M5502JMA, Mendoza, Argentina\\
$^{2}$Consejo Nacional de Investigaciones Cient\'{\i}ficas y T\'ecnicas, Godoy Cruz 2290, C1425FQB,  Buenos Aires, Argentina\\
}

\date{Accepted XXX. Received YYY; in original form ZZZ}

\pubyear{2021}

\begin{document}
\label{firstpage}
\pagerange{\pageref{firstpage}--\pageref{lastpage}}
\maketitle

\begin{abstract}
Globular clusters are prone to lose stars while  moving around the Milky Way. 
These stars escape the clusters and are distributed throughout extended envelopes 
or tidal tails. However, such extra-tidal structures are not observed in all globular clusters,
and yet there is no structural or dynamical parameters that can predict their
presence or absence. NGC\,6864 is an outer halo globular cluster with reported 
no observed tidal tails. We used Dark Energy Camera (DECam) photometry reaching
 $\sim$ 4 mag
underneath its main sequence turnoff to confidently detect an extra-tidal envelope, 
 and stellar debris spread across the cluster outskirts. These features
emerged once robust field star filtering techniques were applied to the fainter end of
the observed cluster main sequence. NGC\,6864 is associated to the
{\it Gaia}-Enceladus dwarf galaxy, among others 28 globular clusters. 
Up-to-date, nearly 64$\%$ of them have been targeted looking for tidal tails and
most of them have been confirmed to  exhibit tidal tails. Thus, the present
outcomes allow us to speculate on the possibility that {\it Gaia}-Enceladus
globular clusters share a common pattern of mass loss by tidal disruption.
 \end{abstract} 


%%%%%%%%%%%%%%%%%%%%%%%%%%%%%%%%%%%%%%%%%%%%%%%%%%

%%%%%%%%%%%%%%%%% BODY OF PAPER %%%%%%%%%%%%%%%%%

\section{Introduction}

Globular cluster tidal tails have become sensitive tracers of the nature and distribution 
of dark matter in the Milky Way \citep{bonacaetal2019}. There are  indications from 
N-body simulations that globular clusters might originally have been embedded inside
substantial mini-halos of dark matter \citep{penarrubiaetal2017,boldrinietal2020}.
Nevertheless, there is an ongoing debate as to whether the gaps observed in streams,
as for example, along the GD-1 stream and the Pal\,5 tidal tails 
\citep{gd2006a,gd2006b}, are due to dark matter sub-halos, or to epicyclic motions of 
a continuous stream of stars escaping the clusters as a  consequence of globular
cluster tidal interaction with the Milky Way \citep{kupperetal2010,kupperetal2012}.
From an observational point of view, there have been many investigations of the 
outermost regions of globular clusters with the goal of detecting extra-tidal structures 
and tidal tails. From a recent comprehensive compilation of the relevant observational 
results obtained to date of 53 globular clusters, \citet{pcb2020}
found that 14 globular clusters have observed tidal tails and 17 show no detectable
signatures of extra-tidal structures. They showed that there is no currently known 
parameters that enable us to confidently predict the presence or absence of tidal tails 
for any given globular cluster.

NGC\,6864 (M75) is a Milky Way globular cluster located at a Galactocentric distance
$R_{GC}$ of 14.7 kpc \citep[][December 2010 Edition]{harris1996}. In the context of that 
a fraction of the Milky Way globular cluster population might have originated in and have
been accreted with already extinct dwarf galaxies, \citet{carballobelloetal2014} 
searched for the remnants of the progenitor galaxy that might be still populating 
the surroundings of this cluster. From cross-correlation and isochrone-fitting methods,
they found no evidence of distinct stellar population concentrated at a specific 
distance within the probed colour-magnitude range. The lack of tidal debris led them
to wonder whether they could rule out an accretion origin for this globular cluster.
They posed that either the progenitor galaxy was not massive or that the accretion
event took place early in the Milky Way hierarchical assembly \citep{bonacaetal2021}.

\citet{deboeretal2019} took advantage of {\it Gaia} DR2 data \citep{gaiaetal2018b}
to study the outskirts of NGC\,6864, using the {\it Gaia} proper motions to select
cluster's members. They assumed that cluster's stars distributed in the
periphery of the cluster share the mean cluster's proper motion and limited
their star sample to those with $G$ $<$ 20 mag. This means that they made use
of red giant stars that move similarly to the cluster's body. However, since
these stars are either in the process of escaping the cluster or orbiting within a dark matter 
mini-halo, their velocity behavior should be somewhat different from that of stars 
still bound to the cluster \citep{starkmanetal2019}, so that proper motions by 
themselves are not sufficient  for reliable membership determination of tidal tails' stars. 
On the other hand, 
less massive main sequence stars are prone to leave the cluster more easily than 
red giant ones. Detected tidal tails in globular clusters are frequently composed by 
cluster main sequence stars that lie $\sim$ 1-2 mag underneath the main sequence 
turnoff \citep[see, e.g.,][]{jg2010,myeongetal2017,shippetal2018}. The two
selection criteria applied by \citet{deboeretal2019} led them to conclude that the 
cluster does not exhibit tidal tails, which is an expected result according to the
reasons mentioned above.

\begin{figure*}
\includegraphics[width=\textwidth]{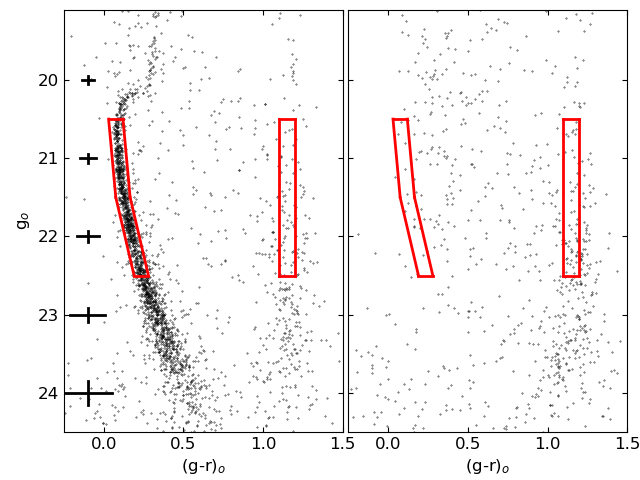}
\caption{Intrinsic CMDs for a circle of radius 0.09$\degr$ centred on the cluster 
(left panel),  compared with that for an annular field region with an equal cluster area
and internal radius of 0.8$\degr$ (right panel). Segments along the cluster main
sequence and the composite field population are drawn, respectively, in red. 
Typical error bars are also indicated.}
\label{fig1}
\end{figure*}

The ex-situ origin of NGC\,6864 was suggested by \citet{massarietal2019}
from the examination of the space of integrals of motion of Milky Way globular clusters,
and confirmed by \citet{forbes2020}, who combined kinematic properties with
ages, metallicities, and alpha-elements, that remain largely unchanged over time.
It belongs to a sample of 28 globular clusters probably associated to the massive
{\it Gaia}-Enceladus galaxy, that underwent a major merger with the 
Milky Way $\sim$8-11 Gyr ago \citep[][and references therein]{naiduetal2021}. 
Nearly 64$\%$ of these globular clusters (18 objects) have studies of their 
outermost regions \citep{pcb2020}, which show that 13 of them (72$\%$) exhibit 
tidal tails or extra-tidal features that are different from tidal tails. A complete census of
{\it Gaia}-Enceladus globular clusters with tidal tails points to the
need of studies of the outskirts of the remaining 10 objects, and of the reanalysis
of those few ones with no previous detection of tidal tails (NGC\,6205, 6229, 6864). 
We note that \citet{jg2010} did not detect any tidal tails in the 
{\it Gaia}-Enceladus globular cluster NGC\,6341, which were
recently uncovered by \citet{thomasetal2020}. This example illustrates that the
combination of different photometry depths and filtering techniques can
produce meaningful distinct results.

In this work we performed a novel search for signatures of tidal tails
around NGC\,6864. Previous studies indicate the lack of observational evidence
of such extra-tidal features, which somehow is counterintuitive for an accreted
outer halo globular cluster. In Sections 2 and 3 we describe the data used and
the developed method applied to highlight the outer faint cluster stellar structures,
respectively. Section 4 deals with the analysis of the resulting stellar density maps.

\begin{figure}
\includegraphics[width=\columnwidth]{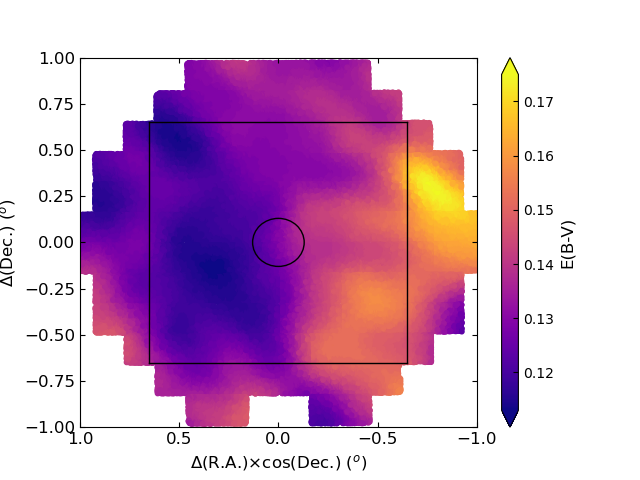}
\caption{Reddening variation across the field of NGC\,6864. The solid box 
delimites the internal boundaries of the adopted field star reference field. The circle 
corresponds to the cluster tidal radius (see Section 4).}
\label{fig2}
\end{figure}

\section{Collected data}

The Dark Energy Camera (DECam), which is attached to the prime focus of the 
4-m Blanco  telescope at Cerro Tololo Inter-American Observatory (CTIO), was
used to image the field of NGC\,6864. DECam is an array of 62 identical chips 
with a scale of 0.263\,arcsec\,pixel$^{-1}$ that provides a 3\,deg$^{2}$ field of view
\citep{flaugheretal2015}. The images, obtained as part of the observing program
CTIO 2019B-1003 (PI : Carballo-Bello), are publicly available and consist of
4$\times$600 sec $g$ and 4$\times$400 sec $r$ exposures, respectively. 
NGC\,7089 is the other globular cluster included in the program 
with unpublished DECam data; it will be the subject of a forthcoming paper.
The data set includes nightly observations of  5 SDSS fields at  a different airmass,
which were used to derive the atmospheric extinction coefficients  and the 
transformations between the instrumental magnitudes and the SDSS $ugriz$ 
system \citep{fukugitaetal1996}.

In order to process the program images, the DECam Community Pipeline 
\citep{valdesetal2014} was used, while the photometry was obtained from the 
processed images using the  \textsc{daophot\,ii/allstar} point-spread-function
fitting routines \citep{setal90}. From the resulting photometric catalog, we 
imposed the restriction of $|$sharpness$|$ $\leq$ 0.5 to avoid the presence of bad 
pixels, cosmic rays, background galaxies, and unrecognized double stars in our 
subsequent analysis. and kept positions and standardized $g$ and $r$ magnitudes 
of stellar objects that surpassed that criterion. The photometry completeness was
estimated from artificial star tests, using the {\sc daophot\,ii} routines. Briefly,
synthetic stars with magnitudes and positions distributed similarly to those of the 
measured stars in each image were added, and their photometry was
carried out applying the same point-spread-function fitting routines, as above.
Finally, the resulting magnitudes for the synthetic stars were then compared to 
those used to create such stars. The 50$\%$ completeness level resulted to be
 23.4\,mag and 23.3\,mag for the $g$ and $r$ bands, respectively 
 \citep[see, also,][]{piattietal2020,piattietal2021}.
 
 \begin{figure*}
\includegraphics[width=\textwidth]{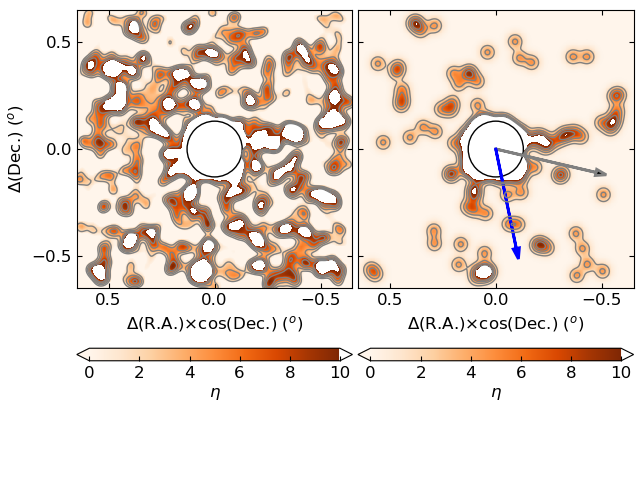}
\caption{Observed cluster main sequence segment stellar density map (left panel) 
and that build for stars that remained unsubstracted from the cleaning procedure
and with $P$ $>$ 70$\%$ (right panel). The black circle centred on the cluster 
indicates the cluster tidal radius (see Section 4). Contours for $\eta$ = 2, 4, 6, and 8 
are also shown. We have painted white stellar densities with $\eta$ $>$ 10 in order 
to highlight the least dense structures. The grey and blue arrows in the right panel
show the direction to the Milky Way centre and that of the motion of the cluster
\citep{baumgardtetal2019},
respectively.}
\label{fig3}
\end{figure*}

Figure~\ref{fig1} (left panel) shows the colour-magnitude diagram (CMD) for a 
circular region around the cluster centre. With the aim of clearly delineate the 
cluster main sequence, we used stars distributed inside a circle with radius 
0.09$\degr$, which is $\sim$ 70$\%$ the cluster tidal radius (see Section 4). 
Thus, we ensured that most of the stars are likely cluster members. As can be seen, 
the cluster main sequence is visibly thin and spans over nearly 4 mag 
underneath its main sequence turnoff. NGC\,6864 is projected along a line-of-sight 
that does not show a severe contamination of field stars. The figure (right panel) 
shows that for an annulus centred on the cluster with an internal radius more 
than 6 times larger than the cluster tidal radius, and with the same cluster
area for comparison purposes, the number of stars along the 
cluster main sequence is notably low. On purpose, the cluster CMD was built
using magnitudes and colours corrected by interstellar reddening, whose
median value across the whole field is low. 
Figure~\ref{fig2} shows the variation of the interstellar reddening $E(B-V)$ 
across the DECam field of view, with $E(B-V)$ values from \citet{sf11} as
provided by the NASA/IPAC Infrared Science Archive\footnote{https://irsa.ipac.caltech.edu/}. 
In order to construct the reddening map, we used a 
uniform grid of Right Ascension and Declination values covering the entire
DECam field of view in steps of 1 arcmin, respectively.
We corrected the $g$ and $r$
magnitudes of each star by using the $E(B-V)$ colour excesses according to the 
positions of the stars in the sky. 

 \begin{figure*}
\includegraphics[width=\textwidth]{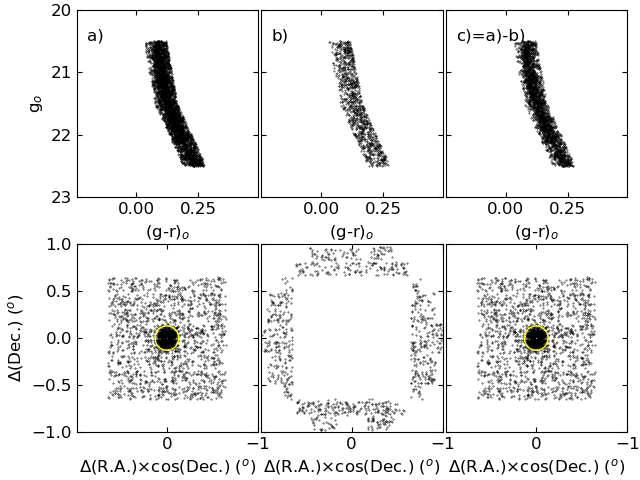}
\caption{{\it Top panels:} CMD of individual stars used in the CMD cleaning 
procedure (see Section 3 for details), namely: a) all the stars in the cleaned area; 
b) all the stars in the
reference field, and c) stars that remained unsubtracted with $P$ $>$ 70$\%$.
{\it Bottom panels:} Spatial distributions of the stars in the  CMDs
built above them, respectively. The yellow circles 
correspond to the cluster tidal radius (see Section 4). }
\label{fig4}
\end{figure*}

\begin{figure}
\includegraphics[width=\columnwidth]{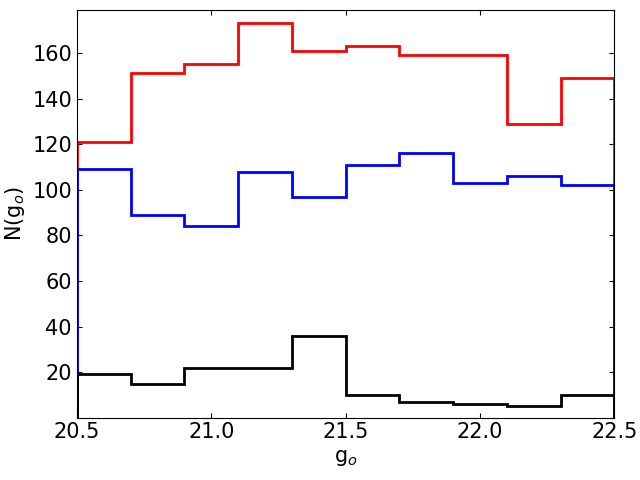}
\caption{$g_0$ distributions in panels a) and c) of Fig.~\ref{fig4}
for stars located outside the cluster tidal radius, drawn with red and
black lines, respective. For comparison purpose, the $g_0$ distribution
of reference field stars is also shown with a blue line.}
\label{fig5}
\end{figure}

\section{Stellar density maps}

\citet{carballobelloetal2014} used a template region along the cluster main sequence
defined from theoretical isochrones and counted the number of stars inside it,
given different weights to the stars according to their distances to the central line
of the template region. Thus, they assigned more weight to stars placed along the
cluster main sequence ridge line. However, such a criterion does not get rid of
field stars, that populate the cluster main sequence too. \citet{deboeretal2019}
employed red giant stars with proper motions similar to the mean cluster proper
motion. We note that mass segregation also makes more massive stars to be more 
centrally concentrated \citep[][and references therein]{khalisietal2007}. We here 
decided to used main sequences stars, because stars with
smaller masses can be found more easily far away from the cluster main body
\citep{carballobelloetal2012}. For this reason, most of the studies devoted to searching 
for extra-tidal structures have used relatively faint main sequence stars 
\citep[see, e.g.,][]{olszewskietal2009,sahaetal2010}. We also cleaned the
cluster main sequence from field star contamination, so that the distribution in the
sky of the stars that remained unsubtracted represents the intrinsic spatial distribution 
of cluster members.

\begin{figure*}
\includegraphics[width=\textwidth]{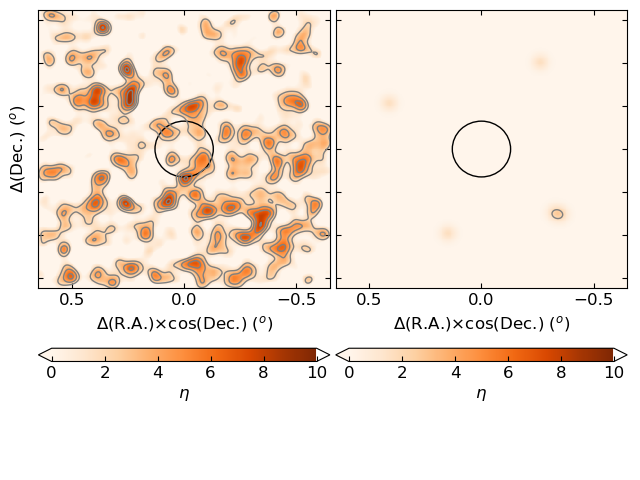}
\caption{Same as Fig.~\ref{fig3} for the control field segment.}
\label{fig6}
\end{figure*}

Figure~\ref{fig1} shows a region on the cluster main sequence enclosed
within red boundaries. We selected that portion of the cluster main sequence
to build the respective stellar density map, once it was cleaned from field star
contamination. The method applied for such a decontamination of field stars
was devised by  \citet{pb12}, and was satisfactorily applied elsewhere 
\citep[e.g.,][and references therein]{p17a,p2018,petal2018}. The procedure
proved to be  successful for clusters projected on to crowded fields and affected by 
differential reddening \citep[see, e.g.,][and references therein]{p17c,pcb2019,pft2020}.
In order to clean the selected cluster main sequence segment, we need
to compare it with a  reference star field, and then to properly eliminate from
the former a number of stars equal to that found in the latter, bearing in mind that 
the magnitudes and colours of the eliminated stars in the cluster main sequence
segment must reproduce the respective magnitude and colour distributions in the 
reference star field main sequence. For that purpose, we traced a square in the
DECam field of view that splits the whole covered field in two equal areas; the
reference star field region being the outermost one.

The methodology to select stars to subtract from the cluster main sequence segment
consists in defining boxes centred on the magnitude and colour of each star of the
reference star field; then to superimpose them on the star cluster main sequence 
segment, and finally to choose one star per box to subtract. We started with
boxes with size of ($\Delta$$g_0$,$\Delta$$(g-r)_0$) = (0.25 mag, 0.10 mag)
centred on the ($g_0$, $(g-r)_0$) values of each reference field star,
in order to guarantee to find a star in the cluster main sequence segment with
the magnitude and colour within the box boundary. In the case that more than one star 
is located inside the cluster main sequence segment, the closest one to the centre of 
that (magnitude, colour) box is subtracted. The magnitude and colour errors of the stars 
in the cluster main sequence segment were taken into account while searching for a 
star to be subtracted. With that purpose, we allowed the stars in the cluster main
sequence segment to vary their magnitudes and colours within an interval of 
$\pm$1$\sigma$, where $\sigma$ represents the errors in their magnitude and colour, 
respectively. We allowed up to 1000 random combination of their magnitude and colour
errors.

Because of the relatively large extension of the cleaned cluster field
(1.3$\degr$$\times$1.3$\degr$; see Fig.~\ref{fig2}), we imposed the condition
that the spatial positions of the stars to be subtracted from the cluster main sequence 
segment were chosen randomly. Thus, we avoided spurious overdensities in the
resulting cluster stellar density map driven by the subtraction of stars from the cluster
main sequence segment that are located nearly in the same sky regions. In practice,
for each reference field star, we first randomly  selected the position of a box of 
0.05$\degr$ a side in the cluster field (the square area in Fig.~\ref{fig2}) where 
to subtract a star. We then looked for a star with ($g_0$, $(g-r)_0$) values within the 
(magnitude, colour) box defined as described above, taking into account the
photometric errors. If no star is found in the selected spatial box, we repeated the 
selection a thousand times, otherwise we enlarged the box size in steps of 
0.01$\degr$ a side, to iterate the process. The outcome of the cleaning procedure
is a cluster main sequence segment that likely contains only cluster members; their 
spatial distributions relies on a random selection. For this reason, we executed 1000 
times the decontamination procedure, and defined a membership probability $P$ 
($\%$) as the ratio $N$/10, where $N$ is the number of times a star was found 
among the 1000 different outputs. In the subsequence analysis we only kept stars 
with $P$ $>$ 70$\%$.
 
The observed stellar density map for stars distributed within the cluster main
sequence segment was built using the \texttt{scikit-learn} software machine learning 
library \citep{scikit-learn} and its kernel density estimator (KDE). We employed a grid 
of 500$\times$500 boxes on the cluster field and allowed the bandwidth to vary from
0.005$\degr$ up to 0.040$\degr$ in steps of 0.005$\degr$. The adopted optimal
value for the bandwidth turned out to be 0.025$\degr$. The background level was 
estimated from the stars distributed in the reference star field.  We split this area in 
boxes of 0.10$\degr$$\times$0.10$\degr$ and counted the number of stars inside 
them, and randomly shifted the boxes by 0.05$\degr$ along the abscisas or ordinates 
and repeated the star counting, with the aim of enlarging the statistics. Finally, 
we derived the  mean  value of the star counts coming from all the defined boxes.
We also estimated its standard deviation from a thousand Monte Carlo realizations 
of the stellar density map, shifting the positions of the stars along 
$\Delta$(RA)$\times$cos(Dec) or $\Delta$(Dec.) randomly (one different shift for 
each star) before recomputing the density map.  Figure~\ref{fig3} depicts the
resulting observed density map (left panel) that shows the absolute deviation 
from the mean value in the field in units of the standard deviation, namely:  
$\eta$ = (signal $-$ mean value)/standard deviation.  The cleaned stellar density
map was constructed using the same recipe and is shown in the right panel
of Fig.~\ref{fig3}. As can be seen, there are several scattered stellar debris around 
the cluster. The direction toward the Milky Way centre
and that of the cluster motion vector\citep{baumgardtetal2019} are also drawn with grey and blue arrows,
respectively.

For the sake of the reader, Fig.~\ref{fig4} illustrates the distribution of stars
in the CMD segment  chosen for building the cluster stellar density map 
(top-left panel); that of the reference star field (top-middle panel), and the
resulting cleaned CMD segment (top-right panel). The bottom panels
show their respective spatial distributions. Stars in the bottom-left and
bottom-right panels were used to build left and right panels of Fig.~\ref{fig3},
respectively. As can be seen, the cleaning procedure subtracted stars 
reproducing the magnitude and colour distributions of stars in the reference
star field and randomly spatially distributed, particularly for stars located
outside the cluster tidal radius. Fig.~\ref{fig5} shows the $g_0$ distributions
of stars in the top-left and top-right panels of Fig.~\ref{fig4} represented by
red and black lines, respectively. For comparison purposes, the 
$g_0$ distribution of stars in the reference star field is also shown with a
blue line. The cleaned cluster CMD segment for stars located
outside the cluster radius shows a slight decrease of the number of stars toward
fainter magnitudes. Such an effect witnesses the cluster mass loss,
being  less massive stars prone to leave the cluster more easily,  
 because they are the first to cross the Jacobi radius once they reach the cluster 
 boundary driven by two-body relaxation.

\section{Analysis and discussion}

Although the cleaned cluster stellar density map was built from stars that have 
chances higher than 70$\%$ of being genuine cluster members, we additionally
validated the field star decontamination procedure by applying it to a sample of 
stars located in a CMD region where only the presence of field stars is expected.
We drew in Fig.~\ref{fig1} a vertical red rectangle spanning the same magnitude 
range as the cluster main sequence segment and placed toward redder colours, 
where field stars are numerous. We note that those stars are located throughout
the entire DECam field of view, as they are seen in both the cluster and
reference star field CMDs (see left and right panels of Fig.~\ref{fig1}).
Therefore, we cleaned the vertical segment composed by stars 
distributed across the cluster field from stars that have magnitudes and colours
within the boundaries of the vertical segment but are located in the reference
star field. Under the assumption that both populations of stars have a common
origin (they belong to the composite Milky Way field population), the 
cleaned stellar density map should show no resulting stellar overdensities. 

We illustrate in Fig.~\ref{fig6} the observed stellar density map built for the
vertical segment (left panel), where stars seem not to be uniformly distributed.
For example,  there are more stars spread in the southwest quadrant than 
toward the north from the cluster centre. We note that vertical segment stars also 
populate  the circle drawn with a radius equal to the cluster tidal radius
and centred on the cluster. The resulting
field star decontaminated stellar density map (right panel of Fig.~\ref{fig6})
shows that there are nearly no residuals with a membership probability 
$P$ $>$ 70$\%$, except an small overdensity located in the south-west
quadrant with $\eta$ $<$ 2. This stellar excess does not coincide with any
extra-tidal features uncovered in Fig.~\ref{fig3} (right panel). Therefore, we
concluded that the uncovered cluster main sequence segment stellar debris 
are intrinsic structural extra-tidal features of NGC\,6864. These findings show
that the combination of appropriate data sets (in this case relatively deep
cluster photometry) and robust field star filtering techniques are suitable
to uncover extra-tidal structures in globular clusters. The stellar
debris detected around NGC\,6864 make it to be added to the number of
outer halo globular clusters with observed extra-tidal features associated to the 
accreted massive {\it Gaia}-Enceladus dwarf galaxy. This finding
confirms that most, if not all, of the globular clusters associated to 
{\it Gaia}-Enceladus exhibit extra-tidal stellar ovendensities 
\citep{carballobelloetal2014,pcb2020,bonacaetal2021}. 
Nevertheless, such an speculative possibility will be comprehensively answered 
once the complete sample of  {\it Gaia}-Enceladus clusters is
targeted by studies that search for  those features around them.

NGC\,6864 shows an excess of stars that extends beyond the cluster tidal
 radius. %and therefore are composed of stars in the process of escaping the cluster. 
\citet{harris1996} compiled a tidal radius of 0.09$\degr$, while \citet{carballobelloetal2012}
estimated a value of 0.11$\degr$. \citet{deboeretal2019} fitted \citet{king1966} and 
\citet{wilson1975} models, a lowered isothermal model explorer in {\sc PYTHON}
\citep[LIMEPY;][]{gm2015}, and a spherical potential escapers stitched model 
\citep[SPES,][]{claydonetal2019} to the number density profile constructed from
cluster red giants. They concluded that both King and Wilson models are too simplistic, 
and {\sc LIMEPY} or SPES models are needed to explain the distribution of stars 
simultaneously in the inner and outer regions. While inspecting their results
(see their figure A13), we found that the best solution is that of the SPES model,
whose resulting tidal radius is 0.13$\degr$. We would like to note that
the different tidal radii derived above, and hence the name of extra-tidal stars
given to those stars located beyond that distance from the custer centre, are
different from the Jacobi radius estimated by \citet[][0.46$\degr$]{bg2018}.  
According to \citet{deboeretal2019}, their SPES tidal radii are similar in general
to the Jacobi radii for Milky Way globular clusters, although their Figure 10 reveals some
noticeable differences.

\begin{figure}
\includegraphics[width=\columnwidth]{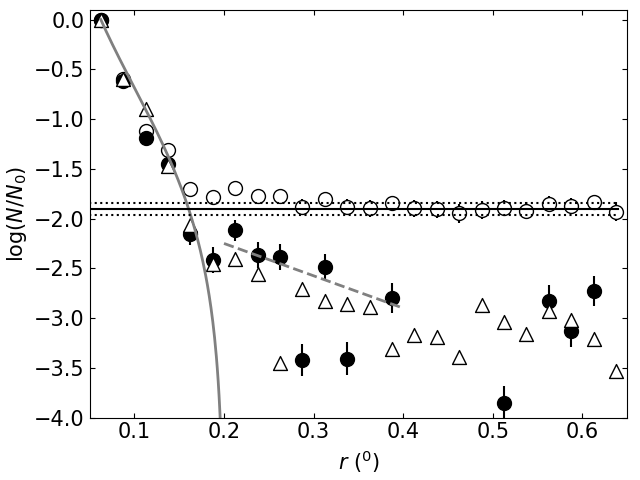}
\caption{ Normalized observed (open circle), mean background subtracted (filled circle) 
stellar radial profiles and that for stars with $P$ $>$ 70 (filled circle). The solid and
dotted horizontal lines represent the mean background level and its dispersion, respectively.
The grey solid line represents the best-fitted \citet{king62}'s profile, while the dashed line corresponds to  a power law with $\alpha$= 3.25.}
\label{fig7}
\end{figure}

We  built the cluster stellar radial profile using all the 
stars distributed in the stellar density map of Fig.~\ref{fig3}. 
We focused on the outermost region ($r$ $>$ 0.09$\degr$) where radial variations 
of the photometry completeness are negligible and where we are interested in 
finding cluster extra-tidal features. In order to generate the stellar radial profile, 
we counted the number of stars in annuli of 0.025$\degr$ wide.
Figure~\ref{fig7} depicts the resulting  observed radial profile represented with open 
circles. We then estimated the mean background level using all the points located at 
distances larger than 0.65$\degr$ from the cluster centre, which turned out to be 
log($N_{bg}/N_0$) = -1.90$\pm$ 0.06. Once the mean background level was subtracted 
from the
observed radial profile, we obtained the radial profile represented by filled circles in
Fig.~\ref{fig7}. For comparison purposes we superimposed the field star cleaned
radial profile constructed from counts of stars found in the decontaminated cluster CMD
with $P$ $>$ 70$\%$ (open trialgles). We fitted the normalized field star cleaned
radial profile with a \citet{king62}'s model using the 
known cluster core radius \citep[0.0015$\degr$;][]{harris1996,baumgardtetal2019}.
The derived tidal radius turned out to be (0.20$\pm$0.04)$\degr$. As can be seen
in Fig.~\ref{fig7}, there is an excess of stars beyond the fitted \citet{king62} profile that
follows a power law  $\propto$ $r^{-\alpha}$, with slope $\alpha$ equals to 3.25. 
This value is in  between those found in Milky Way globular clusters
with extended halo-like structures, e.g., NGC\,1851 and 47\,Tuc 
\citep[][$\alpha$ = 1.24]{olszewskietal2009,p17c} and the abrupt fall
of the $r^{-4}$ law suggested by \citet{penarrubiaetal2017}
as a prediction of expected stellar envelopes of Milky Way globular clusters
 embedded in dark mini-haloes.

%we used the KDE with a grid of 300 points covering distances from the cluster centre
%from 0.09$\degr$ up to 0.65$\degr$, a Gaussian kernel, and a bandwidth of 0.025$\degr$.
%Fig.~\ref{fig5} depicts the resulting radial profile, coloured according to the  
%the absolute deviation from the mean value in the field in units of the standard deviation
%($\eta$). For comparison purposes, we superimposed a \citet{king62} profile for a
%tidal radius of 0.13$\degr$ \citep{deboeretal2019}. As can be seen, excesses of stars
%higher than three times the standard deviation ($\eta$ = 3) are found out to 0.16$\degr$,
%and for $\eta$ = 1  out to 0.58$\degr$.

%We adopted this latter tidal radius
%(see Figs.~\ref{fig2}-\ref{fig4}), which in turn resulted to be the largest value
%estimated for the cluster so far. By using this tidal radius, it is readily visible from
%Fig.~\ref{fig3} the existence of an extra-tidal halo, that covers the whole range
%of position angles. 

In general, globular clusters whose orbits are relatively more eccentric and very inclined 
with respect to the Milky Way plane, have undergone a larger amount of mass loss by tidal 
disruption \citep[see, Fig. 1 of][]{pcb2020}. The inclination of the orbit and eccentricity of
NGC\,6864 are $i$ = 49.76$\degr$ and $\epsilon$= 0.79, respectively \citep{piatti2019}.
However, the cluster has lost 16$\%$ of its initial mass from tidal disruption, which is
a relative low value as compared with the values for other globular clusters with
similar inclinations of their orbits and eccentricities \citep[$>$ 30$\%$;][]{piattietal2019b}.
NGC\,6864 is at 14.7 kpc from the Galactic centre and its peri- and apogalactic
distances are 2.1 and 18.00 kpc, respectively \citep{baumgardtetal2019}. Because of it
varying galactocentric distance, the cluster's size changes, reaching the biggest extension
at its apogalactic distance. For this reason, extra-tidal features could be embraced by
the dynamically enlarged cluster volume and therefore could not be observed. This
does not seem to be the case of NGC\,6864.
Likewise, perturbations along the globular clusters' 
tidal tails such as density variations and changes in their directions have been
observed  \citep{bonacaetal2019,bonacaetal2020}. Unfortunately, the DECam field of
 view is not large enough as to detect tidal tails.

The observed stellar excesses around NGC\,6864 (within a field of 1.3$\degr$$\times$1.3$\degr$) 
and those found in most of the globular clusters associated to 
{\it Gaia}-Enceladus with studies of their outermost regions can be
interpreted to the light of the theory of kinematically chaotic and non-chaotic 
systems \citep[see, e.g.][]{pricewhelanetal2016a,pricewhelanetal2016b,perezvillegasetal2018}. 
We note that
not every Milky Way globular cluster presents tidal tails and that \citet{pcb2020}
using a comprehensive compilation of globular clusters,
did not find any clues as to understand why tidal tails are not seen in every
globular clusters. They explored different structural and kinematical properties and
found that globular clusters behave similarly, independently of the presence of tidal 
tails or any other kind of extra-tidal feature, or the absence thereof. 
Recently, \citet{mestreetal2020} compared the behavior of simulated streams embedded
in chaotic and non-chaotic regions of the phase-space. They find that typical 
gravitational potentials of host galaxies can sustain chaotic orbits, which in turn
do reduce the time interval during which streams can be detected. Therefore, 
tidal tails in some globular clusters are washed out afterwards they
are generated to the point at which it is imposible to detect them. This is the
case of innermost globular clusters, with apogalactic distances smaller than
$\sim$ 6 kpc \citep{pcb2020}. Conversely, outer halo globular clusters spend most
of their lifetimes moving along non-chaotic orbits, so that their tidal tails remain
for a longer period before they are eventually wipe out.

Some recent results suggest that the absence of tidal tails in some globular clusters
is related to the fact that they were born within dark matter sub-halos, and role
reversal \citep{starkmanetal2019,bv2021,wanetal2021}. According to \citet{boldrinietal2020}
halo globular clusters formed at or near the centre of small dark matter halos 
and at present day still retain an excess of dark matter above the galactic background
dark matter. However, NGC\,6205 and NGC\,7099, two {\it Gaia}-Enceladus globular
clusters, do not have observed tidal tails \citep{jg2010,piattietal2020} and only
NGC~6205 shows some evidence of being embedded in a dark matter sub-halo. 
Furthermore, we matched 25 globular clusters, among them NGC~6205, 7099
and other 8 {\it Gaia}-Enceladus globular clusters, tested by \citet{cg2021} while
searching for signatures of being  contained within dark matter sub-halos, 
with the comprehensive compilation of detected tidal tails \citep[see Table 1 of][]{pcb2020} 
and found that several globular clusters without tidal tails do not show a rising 
velocity dispersion profiles as expected for those embedded in dark matter
sub-halos. Therefore, we cannot rule out  considering non-chaotic or chaotic orbits,  
neither globular clusters formed in  dark matter sub-halos to interpret the
existence of tidal tails around globular clusters.

\section{Data availability}

DECam images used in this work are publicly available at the 
https://astroarchive.noao.edu/portal/search/\#/search-form webpage.

\section*{Acknowledgements}
I thank the referee for the thorough reading of the manuscript and
timely suggestions to improve it. 
I also thank the contribution of J.A. Carballo-Bello to an earlier stage of this
project.

Based on observations at Cerro Tololo Inter-American Observatory, NSF’s NOIRLab (Prop. ID 2019B-1003; 
PI: Carballo-Bello), which is managed by the Association of Universities for Research in Astronomy (AURA)
under a cooperative agreement with the National Science Foundation.

This project used data obtained with the Dark Energy Camera (DECam), which was constructed by the 
Dark Energy Survey (DES) collaboration. Funding for the DES Projects has been provided by the US 
Department of Energy, the US National Science Foundation, the Ministry of Science and Education of Spain, 
the Science and Technology Facilities Council of the United Kingdom, the Higher Education Funding Council 
for England, the National centre for Supercomputing Applications at the University of Illinois at 
Urbana-Champaign, the Kavli Institute for Cosmological Physics at the University of Chicago, centre for 
Cosmology and Astro-Particle Physics at the Ohio State University, the Mitchell Institute for Fundamental 
Physics and Astronomy at Texas A\&M University, Financiadora de Estudos e Projetos, Funda\c{c}\~{a}o 
Carlos Chagas Filho de Amparo \`{a} Pesquisa do Estado do Rio de Janeiro, Conselho Nacional de 
Desenvolvimento Cient\'{\i}fico e Tecnol\'ogico and the Minist\'erio da Ci\^{e}ncia, Tecnologia e Inova\c{c}\~{a}o, the Deutsche Forschungsgemeinschaft and the Collaborating Institutions in the Dark Energy Survey.
The Collaborating Institutions are Argonne National Laboratory, the University of California at Santa Cruz, the University of Cambridge, Centro de Investigaciones En\'ergeticas, Medioambientales y Tecnol\'ogicas–Madrid, the University of Chicago, University College London, the DES-Brazil Consortium, the University of Edinburgh, the Eidgen\"{o}ssische Technische Hochschule (ETH) Z\"{u}rich, Fermi National Accelerator Laboratory, the University of Illinois at Urbana-Champaign, the Institut de Ci\`{e}ncies de l’Espai (IEEC/CSIC), the Institut de F\'{\i}sica d’Altes Energies, Lawrence Berkeley National Laboratory, the Ludwig-Maximilians Universit\"{a}t M\"{u}nchen and the associated Excellence Cluster Universe, the University of Michigan, NSF’s NOIRLab, the University of Nottingham, the Ohio State University, the OzDES Membership Consortium, the University of Pennsylvania, the University of Portsmouth, SLAC National Accelerator Laboratory, Stanford University, the University of Sussex, and Texas A\&M University.



%\bibliographystyle{mnras}
%\bibliography{paper} % if your bibtex file is called paper.bib

\begin{thebibliography}{}
\makeatletter
\relax
\def\mn@urlcharsother{\let\do\@makeother \do\$\do\&\do\#\do\^\do\_\do\%\do\~}
\def\mn@doi{\begingroup\mn@urlcharsother \@ifnextchar [ {\mn@doi@}
  {\mn@doi@[]}}
\def\mn@doi@[#1]#2{\def\@tempa{#1}\ifx\@tempa\@empty \href
  {http://dx.doi.org/#2} {doi:#2}\else \href {http://dx.doi.org/#2} {#1}\fi
  \endgroup}
\def\mn@eprint#1#2{\mn@eprint@#1:#2::\@nil}
\def\mn@eprint@arXiv#1{\href {http://arxiv.org/abs/#1} {{\tt arXiv:#1}}}
\def\mn@eprint@dblp#1{\href {http://dblp.uni-trier.de/rec/bibtex/#1.xml}
  {dblp:#1}}
\def\mn@eprint@#1:#2:#3:#4\@nil{\def\@tempa {#1}\def\@tempb {#2}\def\@tempc
  {#3}\ifx \@tempc \@empty \let \@tempc \@tempb \let \@tempb \@tempa \fi \ifx
  \@tempb \@empty \def\@tempb {arXiv}\fi \@ifundefined
  {mn@eprint@\@tempb}{\@tempb:\@tempc}{\expandafter \expandafter \csname
  mn@eprint@\@tempb\endcsname \expandafter{\@tempc}}}

\bibitem[\protect\citeauthoryear{{Balbinot} \& {Gieles}}{{Balbinot} \&
  {Gieles}}{2018}]{bg2018}
{Balbinot} E.,  {Gieles} M.,  2018, \mn@doi [\mnras] {10.1093/mnras/stx2708},
  \href {http://adsabs.harvard.edu/abs/2018MNRAS.474.2479B} {474, 2479}

\bibitem[\protect\citeauthoryear{{Baumgardt}, {Hilker}, {Sollima}  \&
  {Bellini}}{{Baumgardt} et~al.}{2019}]{baumgardtetal2019}
{Baumgardt} H.,  {Hilker} M.,  {Sollima} A.,   {Bellini} A.,  2019, \mn@doi
  [\mnras] {10.1093/mnras/sty2997}, \href
  {http://adsabs.harvard.edu/abs/2019MNRAS.482.5138B} {482, 5138}

\bibitem[\protect\citeauthoryear{{Boldrini} \& {Vitral}}{{Boldrini} \&
  {Vitral}}{2021}]{bv2021}
{Boldrini} P.,  {Vitral} E.,  2021, arXiv e-prints, \href
  {https://ui.adsabs.harvard.edu/abs/2021arXiv210403635B} {p. arXiv:2104.03635}

\bibitem[\protect\citeauthoryear{{Boldrini}, {Mohayaee}  \& {Silk}}{{Boldrini}
  et~al.}{2020}]{boldrinietal2020}
{Boldrini} P.,  {Mohayaee} R.,   {Silk} J.,  2020, \mn@doi [\mnras]
  {10.1093/mnras/staa011}, \href
  {https://ui.adsabs.harvard.edu/abs/2020MNRAS.492.3169B} {492, 3169}

\bibitem[\protect\citeauthoryear{{Bonaca}, {Hogg}, {Price-Whelan}  \&
  {Conroy}}{{Bonaca} et~al.}{2019}]{bonacaetal2019}
{Bonaca} A.,  {Hogg} D.~W.,  {Price-Whelan} A.~M.,   {Conroy} C.,  2019,
  \mn@doi [\apj] {10.3847/1538-4357/ab2873}, \href
  {https://ui.adsabs.harvard.edu/abs/2019ApJ...880...38B} {880, 38}

\bibitem[\protect\citeauthoryear{{Bonaca} et~al.,}{{Bonaca}
  et~al.}{2020}]{bonacaetal2020}
{Bonaca} A.,  et~al., 2020, \mn@doi [\apj] {10.3847/1538-4357/ab5afe}, \href
  {https://ui.adsabs.harvard.edu/abs/2020ApJ...889...70B} {889, 70}

\bibitem[\protect\citeauthoryear{{Bonaca} et~al.,}{{Bonaca}
  et~al.}{2021}]{bonacaetal2021}
{Bonaca} A.,  et~al., 2021, \mn@doi [\apjl] {10.3847/2041-8213/abeaa9}, \href
  {https://ui.adsabs.harvard.edu/abs/2021ApJ...909L..26B} {909, L26}

\bibitem[\protect\citeauthoryear{{Carballo-Bello}, {Gieles}, {Sollima},
  {Koposov}, {Mart{\'{\i}}nez-Delgado}  \& {Pe{\~n}arrubia}}{{Carballo-Bello}
  et~al.}{2012}]{carballobelloetal2012}
{Carballo-Bello} J.~A.,  {Gieles} M.,  {Sollima} A.,  {Koposov} S.,
  {Mart{\'{\i}}nez-Delgado} D.,   {Pe{\~n}arrubia} J.,  2012, \mn@doi [\mnras]
  {10.1111/j.1365-2966.2011.19663.x}, \href
  {http://adsabs.harvard.edu/abs/2012MNRAS.419...14C} {419, 14}

\bibitem[\protect\citeauthoryear{{Carballo-Bello}, {Sollima},
  {Mart{\'\i}nez-Delgado}, {Pila-D{\'\i}ez}, {Leaman}, {Fliri}, {Mu{\~n}oz}  \&
  {Corral-Santana}}{{Carballo-Bello} et~al.}{2014}]{carballobelloetal2014}
{Carballo-Bello} J.~A.,  {Sollima} A.,  {Mart{\'\i}nez-Delgado} D.,
  {Pila-D{\'\i}ez} B.,  {Leaman} R.,  {Fliri} J.,  {Mu{\~n}oz} R.~R.,
  {Corral-Santana} J.~M.,  2014, \mn@doi [\mnras] {10.1093/mnras/stu1949},
  \href {https://ui.adsabs.harvard.edu/abs/2014MNRAS.445.2971C} {445, 2971}

\bibitem[\protect\citeauthoryear{{Carlberg} \& {Grillmair}}{{Carlberg} \&
  {Grillmair}}{2021}]{cg2021}
{Carlberg} R.~G.,  {Grillmair} C.~J.,  2021, arXiv e-prints, \href
  {https://ui.adsabs.harvard.edu/abs/2021arXiv210600751C} {p. arXiv:2106.00751}

\bibitem[\protect\citeauthoryear{{Claydon}, {Gieles}, {Varri}, {Heggie}  \&
  {Zocchi}}{{Claydon} et~al.}{2019}]{claydonetal2019}
{Claydon} I.,  {Gieles} M.,  {Varri} A.~L.,  {Heggie} D.~C.,   {Zocchi} A.,
  2019, \mn@doi [\mnras] {10.1093/mnras/stz1109}, \href
  {https://ui.adsabs.harvard.edu/abs/2019MNRAS.487..147C} {487, 147}

\bibitem[\protect\citeauthoryear{{Flaugher} et~al.,}{{Flaugher}
  et~al.}{2015}]{flaugheretal2015}
{Flaugher} B.,  et~al., 2015, \mn@doi [\aj] {10.1088/0004-6256/150/5/150},
  \href {http://adsabs.harvard.edu/abs/2015AJ....150..150F} {150, 150}

\bibitem[\protect\citeauthoryear{{Forbes}}{{Forbes}}{2020}]{forbes2020}
{Forbes} D.~A.,  2020, \mn@doi [\mnras] {10.1093/mnras/staa245}, \href
  {https://ui.adsabs.harvard.edu/abs/2020MNRAS.493..847F} {493, 847}

\bibitem[\protect\citeauthoryear{{Fukugita}, {Ichikawa}, {Gunn}, {Doi},
  {Shimasaku}  \& {Schneider}}{{Fukugita} et~al.}{1996}]{fukugitaetal1996}
{Fukugita} M.,  {Ichikawa} T.,  {Gunn} J.~E.,  {Doi} M.,  {Shimasaku} K.,
  {Schneider} D.~P.,  1996, \mn@doi [\aj] {10.1086/117915}, \href
  {https://ui.adsabs.harvard.edu/abs/1996AJ....111.1748F} {111, 1748}

\bibitem[\protect\citeauthoryear{{Gaia Collaboration} et~al.,}{{Gaia
  Collaboration} et~al.}{2018}]{gaiaetal2018b}
{Gaia Collaboration} et~al., 2018, \mn@doi [\aap]
  {10.1051/0004-6361/201833051}, \href
  {http://adsabs.harvard.edu/abs/2018A%26A...616A...1G} {616, A1}

\bibitem[\protect\citeauthoryear{{Gieles} \& {Zocchi}}{{Gieles} \&
  {Zocchi}}{2015}]{gm2015}
{Gieles} M.,  {Zocchi} A.,  2015, \mn@doi [\mnras] {10.1093/mnras/stv1848},
  \href {https://ui.adsabs.harvard.edu/abs/2015MNRAS.454..576G} {454, 576}

\bibitem[\protect\citeauthoryear{{Grillmair} \& {Dionatos}}{{Grillmair} \&
  {Dionatos}}{2006a}]{gd2006a}
{Grillmair} C.~J.,  {Dionatos} O.,  2006a, \mn@doi [\apjl] {10.1086/503744},
  \href {https://ui.adsabs.harvard.edu/abs/2006ApJ...641L..37G} {641, L37}

\bibitem[\protect\citeauthoryear{{Grillmair} \& {Dionatos}}{{Grillmair} \&
  {Dionatos}}{2006b}]{gd2006b}
{Grillmair} C.~J.,  {Dionatos} O.,  2006b, \mn@doi [\apjl] {10.1086/505111},
  \href {https://ui.adsabs.harvard.edu/abs/2006ApJ...643L..17G} {643, L17}

\bibitem[\protect\citeauthoryear{{Harris}}{{Harris}}{1996}]{harris1996}
{Harris} W.~E.,  1996, \mn@doi [\aj] {10.1086/118116}, \href
  {http://adsabs.harvard.edu/abs/1996AJ....112.1487H} {112, 1487}

\bibitem[\protect\citeauthoryear{{Jordi} \& {Grebel}}{{Jordi} \&
  {Grebel}}{2010}]{jg2010}
{Jordi} K.,  {Grebel} E.~K.,  2010, \mn@doi [\aap]
  {10.1051/0004-6361/201014392}, \href
  {https://ui.adsabs.harvard.edu/abs/2010A&A...522A..71J} {522, A71}

\bibitem[\protect\citeauthoryear{{Khalisi}, {Amaro-Seoane}  \&
  {Spurzem}}{{Khalisi} et~al.}{2007}]{khalisietal2007}
{Khalisi} E.,  {Amaro-Seoane} P.,   {Spurzem} R.,  2007, \mn@doi [\mnras]
  {10.1111/j.1365-2966.2006.11184.x}, \href
  {http://adsabs.harvard.edu/abs/2007MNRAS.374..703K} {374, 703}

\bibitem[\protect\citeauthoryear{{King}}{{King}}{1962}]{king62}
{King} I.,  1962, \mn@doi [\aj] {10.1086/108756}, 67, 471

\bibitem[\protect\citeauthoryear{{King}}{{King}}{1966}]{king1966}
{King} I.~R.,  1966, \mn@doi [\aj] {10.1086/109918}, \href
  {https://ui.adsabs.harvard.edu/abs/1966AJ.....71..276K} {71, 276}

\bibitem[\protect\citeauthoryear{{K{\"u}pper}, {Kroupa}, {Baumgardt}  \&
  {Heggie}}{{K{\"u}pper} et~al.}{2010}]{kupperetal2010}
{K{\"u}pper} A.~H.~W.,  {Kroupa} P.,  {Baumgardt} H.,   {Heggie} D.~C.,  2010,
  \mn@doi [\mnras] {10.1111/j.1365-2966.2009.15690.x}, \href
  {http://adsabs.harvard.edu/abs/2010MNRAS.401..105K} {401, 105}

\bibitem[\protect\citeauthoryear{{K{\"u}pper}, {Lane}  \&
  {Heggie}}{{K{\"u}pper} et~al.}{2012}]{kupperetal2012}
{K{\"u}pper} A. H.~W.,  {Lane} R.~R.,   {Heggie} D.~C.,  2012, \mn@doi [\mnras]
  {10.1111/j.1365-2966.2011.20242.x}, \href
  {https://ui.adsabs.harvard.edu/abs/2012MNRAS.420.2700K} {420, 2700}

\bibitem[\protect\citeauthoryear{{Massari}, {Koppelman}  \& {Helmi}}{{Massari}
  et~al.}{2019}]{massarietal2019}
{Massari} D.,  {Koppelman} H.~H.,   {Helmi} A.,  2019, \mn@doi [\aap]
  {10.1051/0004-6361/201936135}, \href
  {https://ui.adsabs.harvard.edu/abs/2019A&A...630L...4M} {630, L4}

\bibitem[\protect\citeauthoryear{{Mestre}, {Llinares}  \&
  {Carpintero}}{{Mestre} et~al.}{2020}]{mestreetal2020}
{Mestre} M.,  {Llinares} C.,   {Carpintero} D.~D.,  2020, \mn@doi [\mnras]
  {10.1093/mnras/stz3505}, \href
  {https://ui.adsabs.harvard.edu/abs/2020MNRAS.492.4398M} {492, 4398}

\bibitem[\protect\citeauthoryear{{Myeong}, {Jerjen}, {Mackey}  \& {Da
  Costa}}{{Myeong} et~al.}{2017}]{myeongetal2017}
{Myeong} G.~C.,  {Jerjen} H.,  {Mackey} D.,   {Da Costa} G.~S.,  2017, \mn@doi
  [\apjl] {10.3847/2041-8213/aa6fb4}, \href
  {http://adsabs.harvard.edu/abs/2017ApJ...840L..25M} {840, L25}

\bibitem[\protect\citeauthoryear{{Naidu} et~al.,}{{Naidu}
  et~al.}{2021}]{naiduetal2021}
{Naidu} R.~P.,  et~al., 2021, arXiv e-prints, \href
  {https://ui.adsabs.harvard.edu/abs/2021arXiv210303251N} {p. arXiv:2103.03251}

\bibitem[\protect\citeauthoryear{{Olszewski}, {Saha}, {Knezek}, {Subramaniam},
  {de Boer}  \& {Seitzer}}{{Olszewski} et~al.}{2009}]{olszewskietal2009}
{Olszewski} E.~W.,  {Saha} A.,  {Knezek} P.,  {Subramaniam} A.,  {de Boer} T.,
   {Seitzer} P.,  2009, \mn@doi [\aj] {10.1088/0004-6256/138/6/1570}, \href
  {http://adsabs.harvard.edu/abs/2009AJ....138.1570O} {138, 1570}

\bibitem[\protect\citeauthoryear{{Pe{\~n}arrubia}, {Varri}, {Breen}, {Ferguson}
   \& {S{\'a}nchez-Janssen}}{{Pe{\~n}arrubia}
  et~al.}{2017}]{penarrubiaetal2017}
{Pe{\~n}arrubia} J.,  {Varri} A.~L.,  {Breen} P.~G.,  {Ferguson} A. M.~N.,
  {S{\'a}nchez-Janssen} R.,  2017, \mn@doi [\mnras] {10.1093/mnrasl/slx094},
  \href {https://ui.adsabs.harvard.edu/abs/2017MNRAS.471L..31P} {471, L31}

\bibitem[\protect\citeauthoryear{Pedregosa et~al.,}{Pedregosa
  et~al.}{2011}]{scikit-learn}
Pedregosa F.,  et~al., 2011, Journal of Machine Learning Research, 12, 2825

\bibitem[\protect\citeauthoryear{{P{\'e}rez-Villegas}, {Rossi}, {Ortolani},
  {Casotto}, {Barbuy}  \& {Bica}}{{P{\'e}rez-Villegas}
  et~al.}{2018}]{perezvillegasetal2018}
{P{\'e}rez-Villegas} A.,  {Rossi} L.,  {Ortolani} S.,  {Casotto} S.,  {Barbuy}
  B.,   {Bica} E.,  2018, \mn@doi [\pasa] {10.1017/pasa.2018.16}, \href
  {http://adsabs.harvard.edu/abs/2018PASA...35...21P} {35, e021}

\bibitem[\protect\citeauthoryear{{Piatti}}{{Piatti}}{2017a}]{p17a}
{Piatti} A.~E.,  2017a, \mn@doi [\apjl] {10.3847/2041-8213/834/2/L14}, \href
  {http://adsabs.harvard.edu/abs/2017ApJ...834L..14P} {834, L14}

\bibitem[\protect\citeauthoryear{{Piatti}}{{Piatti}}{2017b}]{p17c}
{Piatti} A.~E.,  2017b, \mn@doi [\apjl] {10.3847/2041-8213/aa8773}, \href
  {http://adsabs.harvard.edu/abs/2017ApJ...846L..10P} {846, L10}

\bibitem[\protect\citeauthoryear{{Piatti}}{{Piatti}}{2018}]{p2018}
{Piatti} A.~E.,  2018, \mn@doi [\mnras] {10.1093/mnras/sty773}, \href
  {https://ui.adsabs.harvard.edu/abs/2018MNRAS.477.2164P} {477, 2164}

\bibitem[\protect\citeauthoryear{{Piatti}}{{Piatti}}{2019}]{piatti2019}
{Piatti} A.~E.,  2019, \mn@doi [\apj] {10.3847/1538-4357/ab3574}, \href
  {https://ui.adsabs.harvard.edu/abs/2019ApJ...882...98P} {882, 98}

\bibitem[\protect\citeauthoryear{{Piatti} \& {Bica}}{{Piatti} \&
  {Bica}}{2012}]{pb12}
{Piatti} A.~E.,  {Bica} E.,  2012, \mn@doi [\mnras]
  {10.1111/j.1365-2966.2012.21694.x}, 425, 3085

\bibitem[\protect\citeauthoryear{{Piatti} \& {Carballo-Bello}}{{Piatti} \&
  {Carballo-Bello}}{2019}]{pcb2019}
{Piatti} A.~E.,  {Carballo-Bello} J.~A.,  2019, \mn@doi [\mnras]
  {10.1093/mnras/stz500}, 485, 1029

\bibitem[\protect\citeauthoryear{{Piatti} \& {Carballo-Bello}}{{Piatti} \&
  {Carballo-Bello}}{2020}]{pcb2020}
{Piatti} A.~E.,  {Carballo-Bello} J.~A.,  2020, \mn@doi [\aap]
  {10.1051/0004-6361/202037994}, \href
  {https://ui.adsabs.harvard.edu/abs/2020A&A...637L...2P} {637, L2}

\bibitem[\protect\citeauthoryear{{Piatti} \& {Fern{\'a}ndez-Trincado}}{{Piatti}
  \& {Fern{\'a}ndez-Trincado}}{2020}]{pft2020}
{Piatti} A.~E.,  {Fern{\'a}ndez-Trincado} J.~G.,  2020, \mn@doi [\aap]
  {10.1051/0004-6361/202037439}, \href
  {https://ui.adsabs.harvard.edu/abs/2020A&A...635A..93P} {635, A93}

\bibitem[\protect\citeauthoryear{{Piatti}, {Cole}  \& {Emptage}}{{Piatti}
  et~al.}{2018}]{petal2018}
{Piatti} A.~E.,  {Cole} A.~A.,   {Emptage} B.,  2018, \mn@doi [\mnras]
  {10.1093/mnras/stx2418}, \href
  {http://adsabs.harvard.edu/abs/2018MNRAS.473..105P} {473, 105}

\bibitem[\protect\citeauthoryear{{Piatti}, {Webb}  \& {Carlberg}}{{Piatti}
  et~al.}{2019}]{piattietal2019b}
{Piatti} A.~E.,  {Webb} J.~J.,   {Carlberg} R.~G.,  2019, \mn@doi [\mnras]
  {10.1093/mnras/stz2499}, \href
  {https://ui.adsabs.harvard.edu/abs/2019MNRAS.489.4367P} {489, 4367}

\bibitem[\protect\citeauthoryear{{Piatti}, {Carballo-Bello}, {Mora}, {Cenzano},
  {Navarrete}  \& {Catelan}}{{Piatti} et~al.}{2020}]{piattietal2020}
{Piatti} A.~E.,  {Carballo-Bello} J.~A.,  {Mora} M.~D.,  {Cenzano} C.,
  {Navarrete} C.,   {Catelan} M.,  2020, \mn@doi [\aap]
  {10.1051/0004-6361/202039012}, \href
  {https://ui.adsabs.harvard.edu/abs/2020A&A...643A..15P} {643, A15}

\bibitem[\protect\citeauthoryear{{Piatti}, {Mestre}, {Carballo-Bello},
  {Carpintero}, {Navarrete}, {Mora}  \& {Cenzano}}{{Piatti}
  et~al.}{2021}]{piattietal2021}
{Piatti} A.~E.,  {Mestre} M.~F.,  {Carballo-Bello} J.~A.,  {Carpintero} D.~D.,
  {Navarrete} C.,  {Mora} M.~D.,   {Cenzano} C.,  2021, \mn@doi [\aap]
  {10.1051/0004-6361/202040038}, \href
  {https://ui.adsabs.harvard.edu/abs/2021A&A...646A.176P} {646, A176}

\bibitem[\protect\citeauthoryear{{Price-Whelan}, {Johnston}, {Valluri},
  {Pearson}, {K{\"u}pper}  \& {Hogg}}{{Price-Whelan}
  et~al.}{2016a}]{pricewhelanetal2016a}
{Price-Whelan} A.~M.,  {Johnston} K.~V.,  {Valluri} M.,  {Pearson} S.,
  {K{\"u}pper} A. H.~W.,   {Hogg} D.~W.,  2016a, \mn@doi [\mnras]
  {10.1093/mnras/stv2383}, \href
  {https://ui.adsabs.harvard.edu/abs/2016MNRAS.455.1079P} {455, 1079}

\bibitem[\protect\citeauthoryear{{Price-Whelan}, {Sesar}, {Johnston}  \&
  {Rix}}{{Price-Whelan} et~al.}{2016b}]{pricewhelanetal2016b}
{Price-Whelan} A.~M.,  {Sesar} B.,  {Johnston} K.~V.,   {Rix} H.-W.,  2016b,
  \mn@doi [\apj] {10.3847/0004-637X/824/2/104}, \href
  {https://ui.adsabs.harvard.edu/abs/2016ApJ...824..104P} {824, 104}

\bibitem[\protect\citeauthoryear{{Saha} et~al.,}{{Saha}
  et~al.}{2010}]{sahaetal2010}
{Saha} A.,  et~al., 2010, \mn@doi [\aj] {10.1088/0004-6256/140/6/1719}, \href
  {http://adsabs.harvard.edu/abs/2010AJ....140.1719S} {140, 1719}

\bibitem[\protect\citeauthoryear{{Schlafly} \& {Finkbeiner}}{{Schlafly} \&
  {Finkbeiner}}{2011}]{sf11}
{Schlafly} E.~F.,  {Finkbeiner} D.~P.,  2011, \mn@doi [\apj]
  {10.1088/0004-637X/737/2/103}, 737, 103

\bibitem[\protect\citeauthoryear{{Shipp} et~al.,}{{Shipp}
  et~al.}{2018}]{shippetal2018}
{Shipp} N.,  et~al., 2018, \mn@doi [\apj] {10.3847/1538-4357/aacdab}, \href
  {https://ui.adsabs.harvard.edu/abs/2018ApJ...862..114S} {862, 114}

\bibitem[\protect\citeauthoryear{{Starkman}, {Bovy}  \& {Webb}}{{Starkman}
  et~al.}{2019}]{starkmanetal2019}
{Starkman} N.,  {Bovy} J.,   {Webb} J.,  2019, arXiv e-prints, \href
  {https://ui.adsabs.harvard.edu/abs/2019arXiv190903048S} {p. arXiv:1909.03048}

\bibitem[\protect\citeauthoryear{{Stetson}, {Davis}  \& {Crabtree}}{{Stetson}
  et~al.}{1990}]{setal90}
{Stetson} P.~B.,  {Davis} L.~E.,   {Crabtree} D.~R.,  1990, in {Jacoby} G.~H.,
  ed.,  Astronomical Society of the Pacific Conference Series Vol. 8, CCDs in
  astronomy. pp 289--304

\bibitem[\protect\citeauthoryear{{Thomas} et~al.,}{{Thomas}
  et~al.}{2020}]{thomasetal2020}
{Thomas} G.~F.,  et~al., 2020, \mn@doi [\apj] {10.3847/1538-4357/abb6f7}, \href
  {https://ui.adsabs.harvard.edu/abs/2020ApJ...902...89T} {902, 89}

\bibitem[\protect\citeauthoryear{{Valdes}, {Gruendl}  \& {DES
  Project}}{{Valdes} et~al.}{2014}]{valdesetal2014}
{Valdes} F.,  {Gruendl} R.,   {DES Project} 2014, in {Manset} N.,  {Forshay}
  P.,  eds,  Astronomical Society of the Pacific Conference Series Vol. 485,
  Astronomical Data Analysis Software and Systems XXIII. p.~379

\bibitem[\protect\citeauthoryear{{Wan} et~al.,}{{Wan}
  et~al.}{2021}]{wanetal2021}
{Wan} Z.,  et~al., 2021, \mn@doi [\mnras] {10.1093/mnras/stab306}, \href
  {https://ui.adsabs.harvard.edu/abs/2021MNRAS.502.4513W} {502, 4513}

\bibitem[\protect\citeauthoryear{{Wilson}}{{Wilson}}{1975}]{wilson1975}
{Wilson} C.~P.,  1975, \mn@doi [\aj] {10.1086/111729}, \href
  {https://ui.adsabs.harvard.edu/abs/1975AJ.....80..175W} {80, 175}

\bibitem[\protect\citeauthoryear{{de Boer}, {Gieles}, {Balbinot},
  {H{\'e}nault-Brunet}, {Sollima}, {Watkins}  \& {Claydon}}{{de Boer}
  et~al.}{2019}]{deboeretal2019}
{de Boer} T.~J.~L.,  {Gieles} M.,  {Balbinot} E.,  {H{\'e}nault-Brunet} V.,
  {Sollima} A.,  {Watkins} L.~L.,   {Claydon} I.,  2019, \mn@doi [\mnras]
  {10.1093/mnras/stz651}, \href
  {https://ui.adsabs.harvard.edu/abs/2019MNRAS.485.4906D} {485, 4906}

\makeatother
\end{thebibliography}

%to be uncommented before sending to editor
%\input{paper.bbl}

%\begin{thebibliography}{99}
%\end{thebibliography}

%%%%%%%%%%%%%%%%%%%%%%%%%%%%%%%%%%%%%%%%%%%%%%%%%%
%%%%%%%%%%%%%%%% APPENDICES %%%%%%%%%%%%%%%%%%%%%

%\appendix


%%%%%%%%%%%%%%%%%%%%%%%%%%%%%%%%%%%%%%%%%%%%%%%%%%

% Don't change these lines
\bsp	% typesetting comment
\label{lastpage}
\end{document}